\documentclass[aip,rsi,amsmath,reprint,amssymb]{revtex4-1}

\usepackage{graphicx}
\usepackage{amssymb}
\usepackage{amsmath}

\begin{document}

\title{Role of non-linear effects and standing waves in microwave spectroscopy: Corbino measurements on superconductors and VO$_2$}

\author{Mario Zin{\ss}er}
\affiliation{1.\ Physikalisches Institut, Universit\"at Stuttgart, 70569 Stuttgart, Germany}

\author{Katrin Schlegel}
\affiliation{1.\ Physikalisches Institut, Universit\"at Stuttgart, 70569 Stuttgart, Germany}

\author{Martin Dressel}
\affiliation{1.\ Physikalisches Institut, Universit\"at Stuttgart, 70569 Stuttgart, Germany}

\author{Marc Scheffler}
\affiliation{1.\ Physikalisches Institut, Universit\"at Stuttgart, 70569 Stuttgart, Germany}


\begin{abstract}
Broadband microwave spectroscopy can probe material properties in wide spectral and temperature ranges.
The quality of such measurements crucially depends on the calibration, which also removes from the obtained spectra signatures of standing waves. 
Here we consider cryogenic Corbino-type reflection measurements on superconductors close to the critical temperature. We show that the non-linear sample response, which relates to sample heating, can lead to strong signatures of standing waves even in a well-calibrated Corbino spectrometer.
We demonstrate our findings with microwave measurements as a function of frequency, power, and temperature and for different length of the microwave transmission line.
Finally we note such non-linear effects beyond the case of superconductors by probing a VO$_2$ thin film at the insulator-metal transition.
\end{abstract}

\maketitle

\section{Introduction}
Optical spectroscopy can elucidate the electronic properties of various solids,\cite{Basov2005,Basov2011}
and the case of microwave spectroscopy addresses rather low energy scales.\cite{Scheffler2013} Here superconductors are of particular interest: microwaves can probe the quasiparticle dynamics, the Cooper pair response, and the superconducting energy gap.\cite{Maeda2005,Thiemann2018}

Microwave spectroscopy on superconductors faces two main challenges: firstly, the weak microwave absorption of superconductors demands optimized detection.
Secondly, as the cm-range wavelength at GHz frequencies matches typical dimensions of an experiment, any reflection of the microwaves (at discontinuities of the microwave line) causes pronounced standing waves that in frequency-dependent data show up as Fabry-P\'erot oscillations overlapping the signal of interest.
The standing wave pattern depends on damping and phase shift of the microwave line, and for cryogenic experiments it thus strongly depends on temperature.

Two quite different approaches overcome these challenges:
the first employs a microwave resonator that is perturbed by the sample of interest.
For a low-loss resonator, its bandwidth is much narrower than the characteristic frequencies of the parasitic standing waves of the setup, which then do not disturb detection of resonator frequency and bandwidth.\cite{Klein1993,Petersan1998}
Main disadvantage of such resonant approaches are their limitations in obtaining frequency-dependent information on the sample.\cite{DiIorio1988,Huttema2006,Hafner2014}

If instead one aims at the full frequency dependence, then one has to turn to broadband microwave spectroscopy.\cite{Turner2004,Wiemann2015,Ghirri2015}
Here the most common approach to study superconductors is Corbino reflectometry\cite{Wu1995,Tosoratti2000,Kitano2004,Ohashi2006,Sarti2007,Steinberg2008,Xu2009,Pompeo2010,Liu2011,Liu2013,Mondal2013,Nabeshima2018} where the flat sample terminates a coaxial transmission line, i.e.\ the sample shorts inner and outer conductors and thus reflects the microwave signal that travels on the line.
If the superconducting sample is a very thin film, then its effect on the reflected microwave is strong enough to be detected,\cite{Scheffler2005a} and one can directly determine the frequency-dependent complex conductivity of the sample from the measured complex reflection coefficient $\hat S_{11}$.\cite{Booth1994}
In such an experiment, standing waves cannot be avoided completely, and therefore appropriate calibration is needed, which is very demanding for cryogenic microwave setups.\cite{Booth1994,Scheffler2005a,Stutzman2000,Ohashi2006,Kitano2008,Ranzani2013,Yeh2013}
For Corbino spectrometers, a rigorous calibration involves three different standards (usually open, short, and load) that have to be measured at all frequencies and temperatures of interest.\cite{Scheffler2005a,Stutzman2000,Kitano2008,Liu2014}

As we will show, even a well-calibrated broadband microwave experiment can be prone to standing waves, namely for non-linear effects.
This is the case for superconducting samples close to their critical temperature $T_{\text c}$, when the sample properties depend strongly on temperature. 
While intrinsic non-linear effects in superconductors have been studied previously by microwave experiments (though not in the context of broadband Corbino spectroscopy),\cite{Nguyen1993,Andreone1993,Golosovsky1995,Habib1998,Velichko2005,Kermorvant2009} we here explicitly address non-linear behavior that is related to sample heating.

\begin{table*}[]
	\centering
	\begin{tabular}{ccccccc}
		\hline
		\textbf{\#} & \textbf{setup} & \textbf{VNA} & \textbf{coaxial cables} & 
		$\begin{array}{cc}\boldsymbol{L_\mathrm{phys}}\, \textbf{[cm]}\end{array}$ & $\begin{array}{cc}\boldsymbol{L_\mathrm{el}}\, \textbf{[cm]}\end{array}$ & \textbf{references}\\\toprule
		1 & $\begin{array}{cc}\text{$^4$He bath cryostat}\\\text{(version 1)}\end{array}$ & HP 85107B & UT-085C-TP-LL & 115 & 151 & [\onlinecite{Scheffler2005a}], [\onlinecite{Steinberg2008}]\\
		\hline
		2 & $\begin{array}{cc}\text{$^4$He bath cryostat}\\\text{(version 2)}\end{array}$ & HP 85107B & UT-085B-SS & 114 & 162 & [\onlinecite{Steinberg2008}], [\onlinecite{Steinberg2012}] \\
		\hline
		3 & $^3$He bath cryostat & HP 85107B & $\begin{array}{cc}\text{UT-085C-TP-LL and}\\\text{UT-085B-SS}\end{array}$ & 161 & 236 & [\onlinecite{Steinberg2012}] \\
		\hline
		4 & $\begin{array}{cc}\text{$^4$He flow cryostat}\\\text{(version 1)}\end{array}$ & ENA E5071C & $\begin{array}{cc}\text{flexible cable\cite{MultiplePasternackCableLengths}}\\\text{and UT-085B-SS}\end{array}$ & 45 - 135 & 64 - 194 & $\begin{array}{cc}\text{similar to}\\\text{[\onlinecite{Booth1996b}] and [\onlinecite{Schwartz2000}]}\end{array}$\\
		\hline
		5 &  $\begin{array}{cc}\text{$^4$He flow cryostat}\\\text{(version 2)}\end{array}$ & ENA E5071C & UT-085-SS-SS & 42 & 60 & $\begin{array}{cc}\text{similar to}\\\text{[\onlinecite{Booth1996b}] and [\onlinecite{Schwartz2000}]}\end{array}$\\
		\hline
	\end{tabular}
	\caption{Overview of the different microwave setups used in this study. The coaxial cables are different types of semirigid coax; setup \#4 was used with several different additional flexible room-temperature coaxial cables to change the length of the microwave line.
	The electrical length $L_\mathrm{el}$ of the coaxial cables is calculated from the physical length $L_\mathrm{phys}$ via Eq. (\ref{EqLph}).}
	\label{cryos}
\end{table*}

\section{Experiment}

We perform temperature-dependent Corbino reflectometry measurements, using setups as listed in Table~\ref{cryos}.
Setups \#1, \#2, and \#3 employ a Hewlett-Packard HP 85107B vector network analyzer (VNA) and cover the frequency range from 45~MHz to 40~GHz,\cite{Scheffler2005a} whereas setups \#4 and \#5 use an Agilent Technologies E5071C ENA series VNA for frequencies 300~kHz to 20~GHz.
For all experiments the microwave output power of the VNA was set to a constant value throughout the full measured spectral range.

The calibration of the setups differs as follows: setups \#1 and \#2 ($^4$He bath cryostat) were fully calibrated with three standard samples 
(teflon as open, bulk aluminum as short, NiCr film as load) 
at each temperature of a sequence of measurements, following the established procedure with separate cooldowns.\cite{Stutzman2000, Scheffler2005a}
Setup \#3 ($^3$He bath cryostat) can also be fully calibrated this way,\cite{Steinberg2012} but for experiments as presented below, with numerous finely-space measurements for different temperatures or powers, the presented spectra either just use the room-temperature calibration of the setup or normalization of the spectra of interest to a reference spectrum well above $T_{\text c}$, which is sufficient to present the spectral features of interest in this study. The same holds for setup \#4 ($^4$He flow cryostat; various coaxial cables) whereas its modification, setup \#5 ($^4$He flow cryostat; single semirigid cable), was fully calibrated for all temperatures. 

All these setups allow in-situ measurements of the dc resistance of the Corbino sample as a function of temperature, using the microwave coax as leads.\cite{Scheffler2005a,Scheffler2010} Since this is a two-point measurement, the obtained resistance data contain an offset due to wiring and contact resistance between sample and Corbino probe. This offset usually only weakly depends on temperature and can often be easily subtracted from the data, but all dc data presented below are raw data that still include this offset.

The in-situ dc measurement also allows us to evaluate temperature gradients in the setups, which are substantial in some cases, but if needed can be corrected for with help of reference samples.\cite{Scheffler2005a}
In the work below, we did not correct for this, and thus the temperatures of the presented dc curves, obtained upon cooling down, represent actual sample temperatures, whereas the temperatures listed with the microwave spectra are nominal temperature sensor readings, which, upon heating for constant temperatures during the microwave measurements, can systematically differ from the actual sample temperature. This explains why in several cases below the $T_{\text c}$ of dc data apparently does not match the $T_{\text c}$ relevant for the microwave spectra. Furthermore, the data obtained on NbTiN with setup \#4 have an overall temperature offset.\cite{TemperatureSetup4}

The samples under study are various superconductors (Al, UNi$_2$Al$_3$, NbTiN, Pb) near the superconducting transition and VO$_2$ near the insulator-metal transition. Details of the samples can be found in Appendix~A. Key aspect for the present study is that all these samples feature a strongly temperature-dependent impedance near their respective phase transitions.

\section{Broadband Measurements on Superconducting Thin Films}

\subsection{Phenomenology: Corbino measurements at the superconducting transition}

\begin{figure}
	\centering
	\includegraphics[width=\columnwidth]{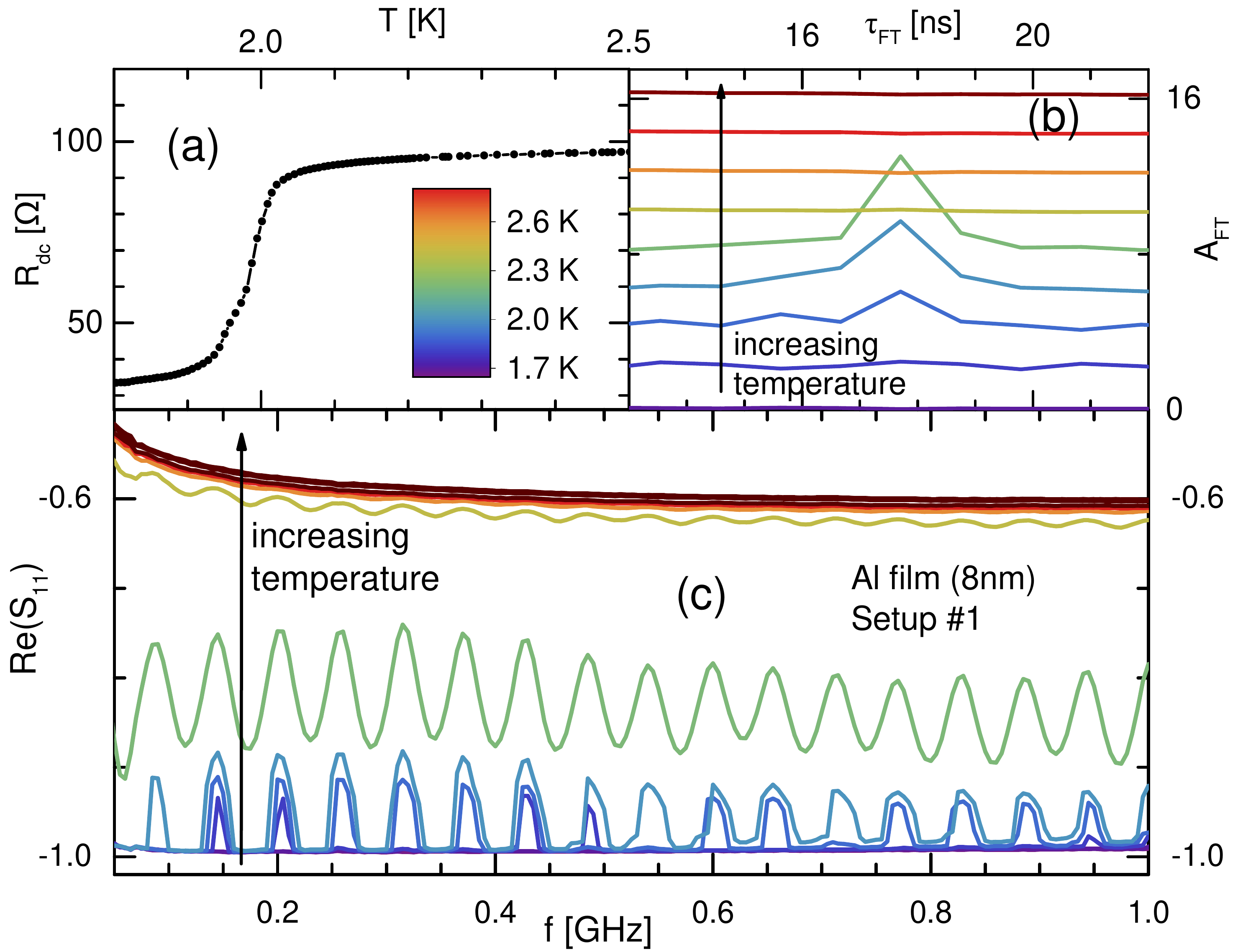}
	\caption{Broadband microwave measurements using fully calibrated Corbino spectrometer (setup \#1) on a superconducting aluminum thin film close to $T_{\text c}$.
		\textbf{(a)} Temperature-dependent $R_\mathrm{dc}$. \textbf{(b)} Fourier transform of Re$(\hat S_{11})$ with stack offset of 2. \textbf{(c)} Microwave spectra of real part Re$(\hat S_{11})$ of reflection coefficient. Close to $T_{\text c}$, pronounced oscillations are visible in the spectra and corresponding peaks in the Fourier transform.}
	\label{Al8nm}
\end{figure}

Fig.\ \ref{Al8nm}(a) shows the temperature-dependent dc resistance $R_\mathrm{dc}$ of an aluminum thin film with clear superconducting transition at $T_{\text c} \approx 2.0$~K.\cite{Scheffler2005a} 
Fig.\ \ref{Al8nm}(c) shows Corbino spectra, in particular the real part Re($\hat S_{11}$) of the reflection coefficient $\hat S_{11}$, obtained on this sample in setup \#1 for several temperatures close to $T_{\text c}$.
Here we focus on rather low frequencies below 1~GHz although we can observe the feature of interest up to much higher frequencies.
For the lowest temperatures, Re$(\hat S_{11})$ is close to $-1.0$, corresponding to a microwave short, as expected for a superconducting sample. 
With rising temperature, Re$(\hat S_{11})$ increases to an almost constant value around $-0.6$ for temperatures above $T_{\text c}$. The rather flat spectra at temperatures well below and well above $T_{\text c}$ demonstrate that the cryogenic three-standard calibration works well for this setup.\cite{Scheffler2005a}
But for temperatures near $T_{\text c}$, the reflection coefficient exhibits pronounced oscillations as a function of frequency.

To characterize those oscillations, a Fourier transform is performed on the Re$(\hat S_{11})$ spectra and shown in Fig.\ \ref{Al8nm}(b).
Here the x-axis quantity we call timefrequency $\tau_{\text{FT}}$, defined as the inverse of any period $p$ observed in the frequency domain, $\tau_{\text{FT}} = \frac 1p$.
Evolving from low to high temperatures, a peak in the Fourier transform at timefrequency $\tau_{\text{osc}}= 17.7$~ns first grows, but then vanishes again, which 
matches the occurrence of the oscillations in the frequency domain in Fig.\ \ref{Al8nm}(c) only near $T_{\text c}$.

\begin{figure}
	\centering
	\includegraphics[width=\columnwidth]{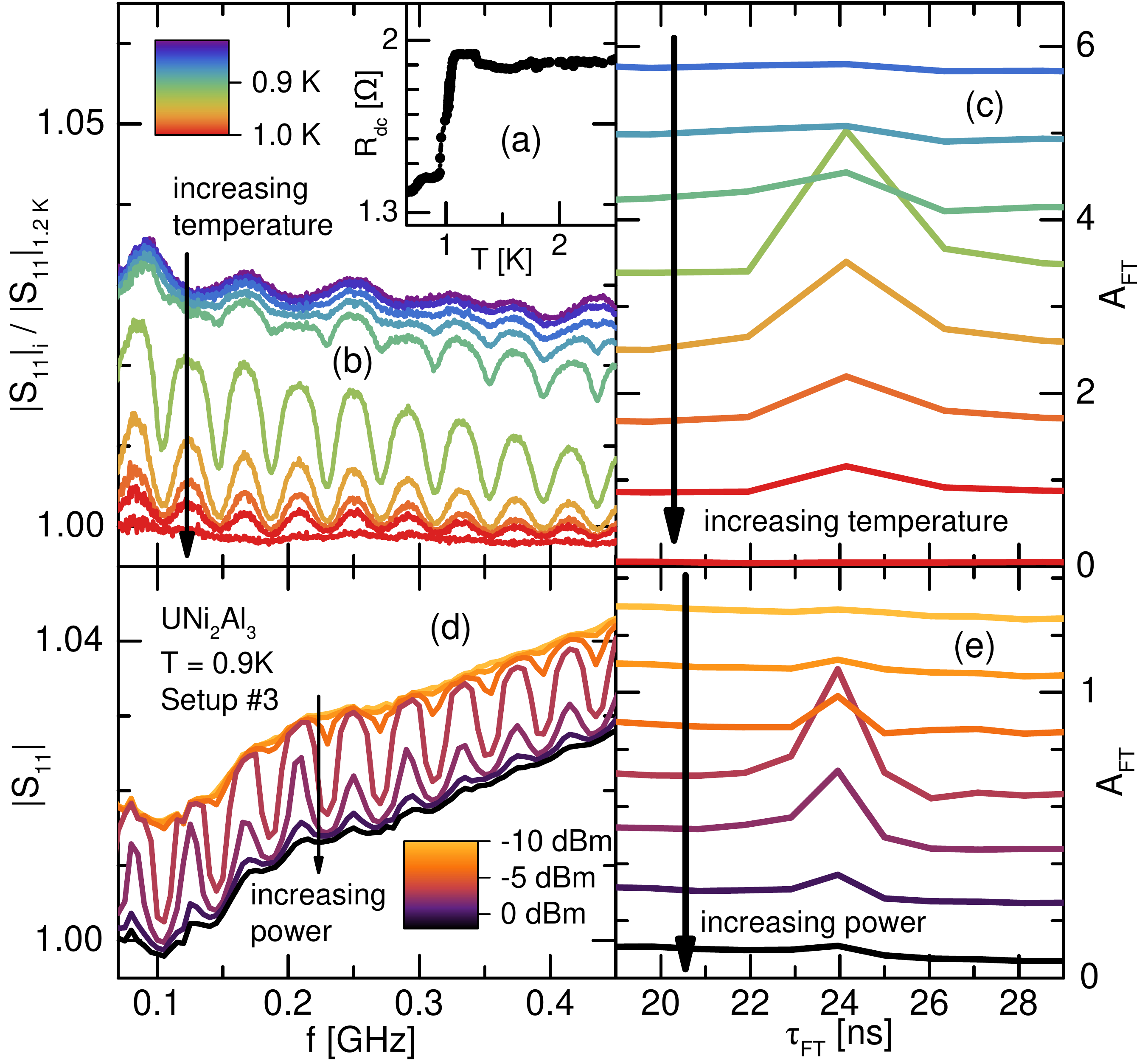}
	\caption{Temperature- and power-dependent microwave measurements on UNi$_2$Al$_3$ 
thin film
near $T_{\text c}$, using setup \#3.
	\textbf{(a)} $R_\mathrm{dc}$ exhibiting $T_{\text c} = 1.0$~K. \textbf{(b)} Spectra of $|\hat S_{11}|$ for several temperatures, normalized to $|\hat S_{11}|$ at 1.2~K. \textbf{(c)} Fourier transform of $|\hat S_{11}|$ spectra in (b), with stack offset of 0.8 and color coding as in (b). \textbf{(d)} $|\hat S_{11}|$ spectra at 0.9~K, obtained for different microwave power. \textbf{(e)} Fourier transform of $|\hat S_{11}|$ spectra in (d), with stack offset of 0.2 and color coding as in (d).}
	\label{UNA}
\end{figure}

In Fig.\ \ref{UNA} we address data of a different superconductor, UNi$_2$Al$_3$ with $T_{\text c} = 1.0$~K evident from the in-situ dc data in Fig.\ \ref{UNA}(a). The microwave spectra were obtained on setup \#3, and the presented data are normalized to the spectrum at 1.2~K. Fig.\ \ref{UNA}(b) shows $|\hat S_{11}|$ spectra for several temperatures, and again we find very strong oscillations in the spectra at the superconducting transition that are absent at temperatures either deep in the superconducting or deep in the normal state.
This oscillation in the spectra corresponds to a peak near $\tau_{\text{osc}}=24$~ns in the Fourier transform in Fig.\ \ref{UNA}(c), which is only present for intermediate temperatures near $T_{\text c}$.
In Fig.\ \ref{UNA}(d) we show several $| \hat S_{11}|$ spectra, all at the same nominal temperature just below $T_{\text c}$ but with different applied microwave power. In case of purely linear response, the measured $\hat S_{11}$ does not depend on signal power, but here it clearly does.
Spectra with high or low powers are rather flat (but differ from each other), whereas intermediate spectra show distinct oscillations, which are again bounded by the outer flat ones. This leads to peaks near $\tau_{\text{osc}}=24$~ns in the Fourier transform in Fig.\ \ref{UNA}(e) only for powers around -5~dBm, but not for substantially higher or lower power.
More detailed data on the interplay of temperature and power are presented in Appendix~B.

\subsection{Cause of oscillations in spectra: sample heating due to standing waves}

Observing regular oscillations in broadband microwave spectra usually calls for consideration of standing waves, but our present phenomenon of the oscillations appearing only near $T_{\text c}$ goes beyond conventional interference in microwave transmission lines. As evident from the spectra well below or well above $T_{\text c}$ in Fig.\ \ref{Al8nm} that do not show oscillations, this setup is well calibrated, i.e.\ the calibration scheme effectively removes signs of standing waves from the raw data. 
Furthermore, some weak oscillations as remnants of a not-perfect calibration in spectra well below $T_{\text c}$ in Fig.\ \ref{UNA}(b) (and Fig.\ \ref{Pb60nm}(b) in Appendix~B) can have a very different oscillation period than the additional, strong oscillation that occurs only near $T_{\text c}$.

As we will demonstrate, the cause for the oscillation near $T_{\text c}$ is the combination of standing waves and non-linear behavior of the sample under study. 
Depending on the standing wave formation in the setup, the microwave power present at the sample position can oscillate as a function of frequency, even though the power submitted by the VNA into the setup is the same for all frequencies and the absorption properties of coaxial cables and sample only vary weakly with frequency. 
Oscillating microwave power at the sample means oscillating power dissipation in the sample, which causes heating of the sample, i.e.\ the actual sample temperature can differ from the nominal temperature measured by the nearby temperature sensor.
Since the impedance of the sample strongly depends on temperature near $T_{\text c}$, the oscillating sample temperature means oscillating sample impedance, which in turn causes an oscillating reflection coefficient, which is the experimental signature that we observe.

\begin{figure}
	\centering
	\includegraphics[width=\columnwidth]{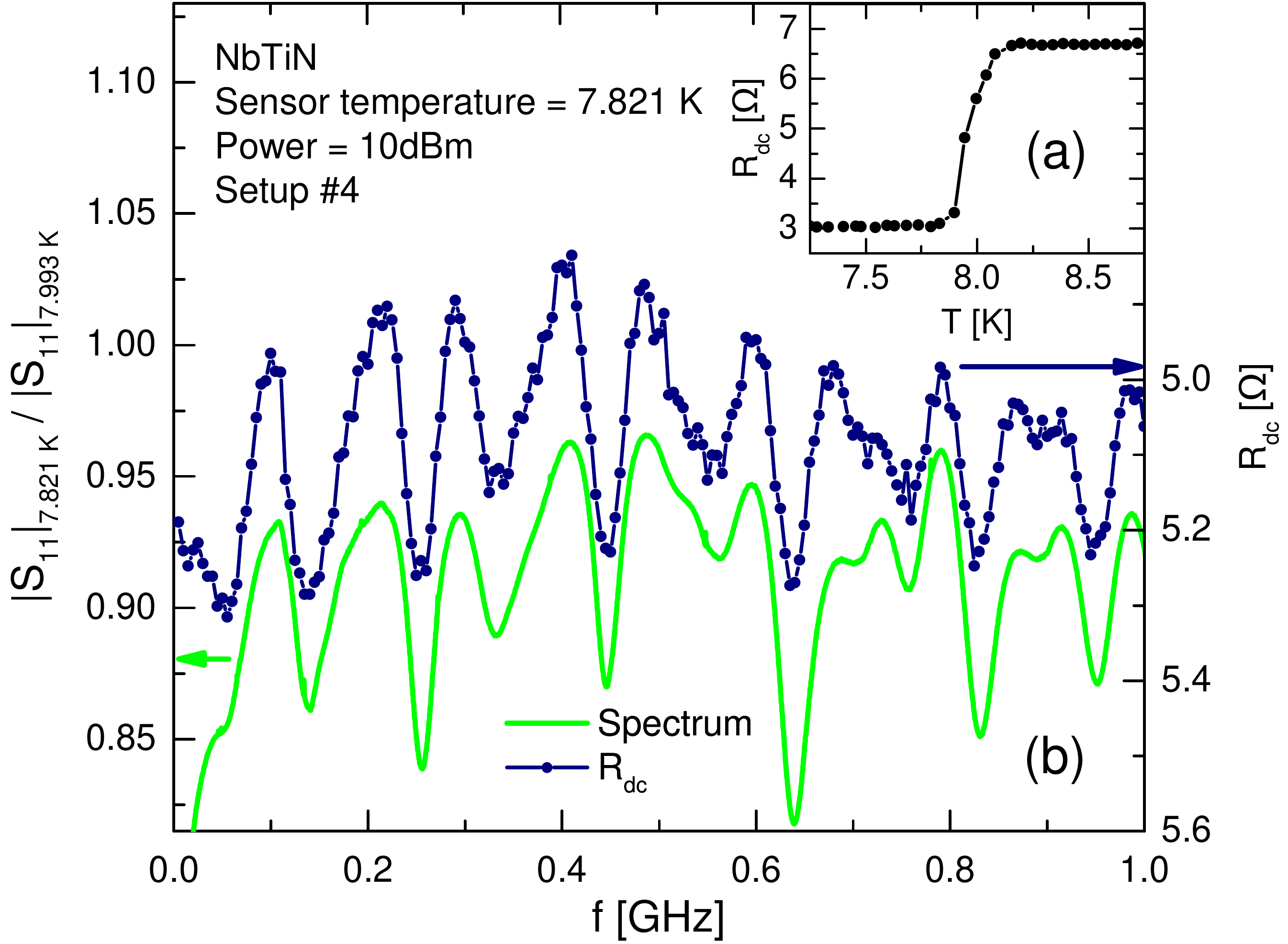}
	\caption{Broadband microwave measurement on NbTiN thin film in setup \#4. \textbf{(a)} Dc resistance $R_\mathrm{dc}$ with superconducting transition near sensor temperature 7.9~K.\cite{TemperatureSetup4} \textbf{(b)} Microwave spectrum obtained at 7.821~K (full line, left axis), with the plotted $|\hat S_{11}|$ normalized to the data obtained at 7.993~K in the metallic state. While the microwave frequency was swept to obtain this spectrum, the two-point $R_\mathrm{dc}$ of the sample was measured simultaneously, and is plotted as full circles (right axis). The oscillations of $R_\mathrm{dc}$ indicate that the sample temperature oscillates, which goes hand in hand with oscillating microwave impedance and thus the oscillating $|\hat S_{11}|$ spectrum.}
	\label{R_DC-Oszis}
\end{figure}

To demonstrate that the oscillations near $T_{\text c}$ indeed are related to temperature changes of the sample, we perform in-situ dc measurements while the microwave frequency is swept. In Fig.\ \ref{R_DC-Oszis} we show such data obtained in setup \#4 on a NbTiN thin film: the microwave spectrum (full line) obtained slightly below $T_{\text c}$ and normalized to a spectrum slightly above $T_{\text c}$ shows pronounced oscillations as discussed before, though with a more complicated frequency dependence than in the previous cases.
The simultaneously measured dc resistance also oscillates substantially (full circles), and the observed oscillatory behavior closely matches the one of the $|\hat S_{11}|$ spectrum, thus confirming that the two are directly related.

Next we evaluate the standing waves in more detail. As mentioned in the introduction, standing waves are a serious challenge for broadband microwave spectroscopy, in particular at cryogenic temperatures. A main goal of the different calibration schemes is to take the standing waves into account, i.e.\ spectra obtained with a properly calibrated setup should not feature standing-wave-caused oscillatory behavior. 
However, if the calibration is imperfect, then the calibrated spectra often contain standing-wave-type oscillations, which do not necessarily reflect the dominant standing waves physically present in the setup, but rather the dominant systematic error of the calibration in terms of standing waves.
In contrast, for the present oscillation observed close to $T_{\text c}$ the cause is heating due to the standing waves physically present at the sample position, i.e.\ it is independent of the calibration scheme. 
A Corbino measurement on a superconducting sample has (at least) two elements that reflect the microwave signal on the line, namely the sample of interest and the VNA. If the coaxial line connecting VNA and Corbino probe does not feature any impedance mismatches, the standing wave pattern is governed by the total electrical length of this transmission line. 

\begin{figure}
	\centering
	\includegraphics[width=\columnwidth]{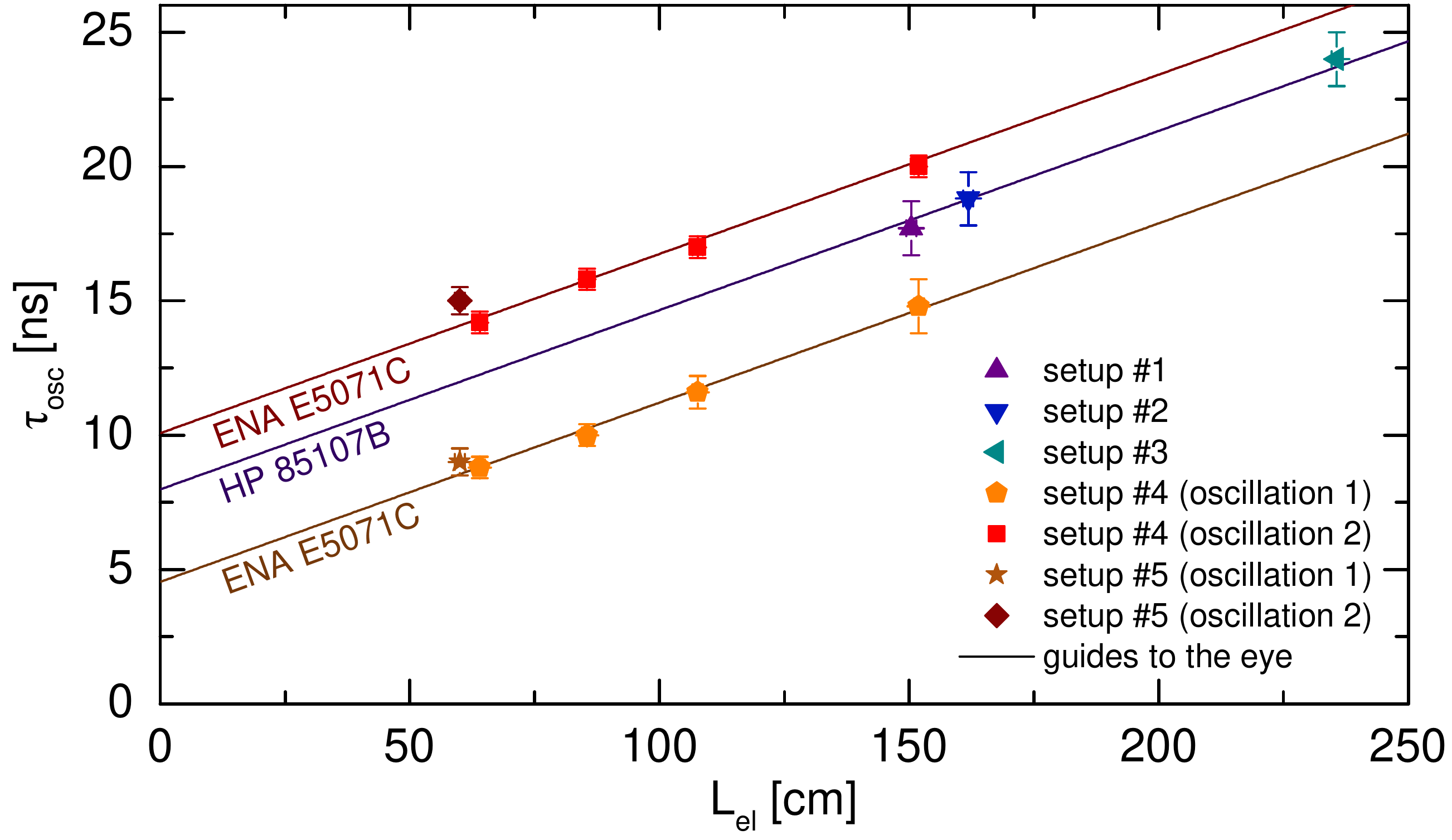}
	\caption{Timefrequencies $\tau_{\text{osc}}$ of the characteristic oscillations for setups with different electrical length $L_\mathrm{el}$ between VNA port and sample. Setup \#4 was measured with several cable lengths. For setups \#4 and \#5, two independent oscillation timefrequencies were found. Vertical error bars indicate width of the peaks in the Fourier transform. The guiding lines indicate the slope expected from Eq.\ \ref{EqfT}, and they are assigned to the data according to the respective VNAs, as labeled.}
	\label{L_eff}
\end{figure}

To test this influence of the length of the coaxial cable, we take advantage of the substantially different cable lengths of the setups listed in Table \ref{cryos}. 
The characteristic timefrequency $\tau_{\text{osc}}$, i.e.\ the time the signal needs for one round trip between the reflecting points on the coaxial line that cause the Fabry-P\'erot-type oscillations, is the inverse of the distance $p_{\text{osc}}$ between the resonance frequencies, $f_k$ and $f_{k+1}$, of two adjacent standing-wave modes (with mode number $k$ and $k+1$) and depends on $L_\mathrm{el}$ as follows (with $c_0$ the speed of light in vacuum):
\begin{equation}
\tau_{\text{osc}} = \frac{1}{p_{\text{osc}}}=  \frac{1}{f_{k+1} - f_k} = \frac{1}{\frac{(k+1) \, c_0}{2L_\mathrm{el}} - \frac{k \, c_0}{2L_\mathrm{el}}} = \frac{2}{c_0}L_\mathrm{el}
\label{EqfT}
\end{equation}
The electrical length $L_\mathrm{el}$ of the coaxial line depends on its physical length $L_\mathrm{phys}$ and its effective dielectric constant $\epsilon_\mathrm{eff}$:
\begin{equation}
L_\mathrm{el} = \sqrt{\epsilon_\mathrm{eff}} \, L_\mathrm{phys}
\label{EqLph}
\end{equation}
The values of $L_\mathrm{phys}$ in Table \ref{cryos} are the measured lengths of the coaxial cables between VNA port and Corbino sample, and for $\epsilon_\mathrm{eff}$ we turned to the specifications of the cable manufacturers.\cite{fToffset}

In Fig.\ \ref{L_eff} the timefrequencies $\tau_{\text{osc}}$ of the Fourier transform peaks are plotted vs.\ $L_\mathrm{el}$ for all setups of this study. To extend the range of cable lengths beyond the dedicated low-temperature setups of our lab, we modified setup \#4 by inserting additional flexible cables into the room-temperature section of its coaxial line, and then performed measurements on the NbTiN sample near $T_{\text c}$ to evaluate the characteristic oscillations in the spectra. 
For the measurements with setups \#4 and \#5, which share the same VNA and cryostat, we observed two distinct timefrequencies, and both are plotted in Fig.\ \ref{L_eff}. (These two separate oscillation frequencies are also the reason for the \lq beating\rq{} pattern visible in Fig.\ \ref{R_DC-Oszis}.)
Based on Eq.\ \ref{EqLph}, we expect linear behavior for Fig.\ \ref{L_eff} with slope $2 / c_0$, but with an unknown offset because our $L_\mathrm{phys}$ does not contain the part of the transmission line within the VNA. Indeed we find that the data can be classified as three groups with different offsets (setups \#1 to \#3 with HP 85107B VNA, first oscillations of setups \#4 and \#5 with ENA E5071C VNA, and their second oscillations, respectively) but all with the expected slope $2 / c_0$, as indicated by the straight lines in Fig.\ \ref{L_eff}.
These data also signal how the VNA hardware can affect the oscillations.

These experiments confirm the interpretation that the pronounced oscillations in the Corbino spectra near $T_\mathrm{c}$ are caused by the combination of standing waves on the coaxial line and pronounced non-linear response of the sample in the vicinity of the superconducting transition. 
This also explains the very \lq inharmonic\rq{} oscillation spectra that we have observed in several cases e.g.\ in Fig.\ \ref{Al8nm}(c)-(e) (blue curves), where the sample resides in the superconducting state for certain frequencies with low local microwave power, but if the standing waves induce a local microwave power that is strong enough to cause a temperature increase of the sample into the steep section of the $R_\mathrm{dc}(T)$ curve, then the sample impedance will increase drastically. 
The observed oscillations in the $\hat S_{11}$ spectra thus stem from oscillations of the sample impedance, but this is not an intrinsic frequency dependence of the sample but a measurement artifact caused by the oscillating sample temperature caused by the physical standing waves at the sample.

\section{Implications for Corbino Spectroscopy}

We have shown that broadband Corbino measurements are prone to oscillatory artifacts in the microwave spectra that stem from standing waves combined with non-linear sample response due to sample heating. 
This phenomenon can easily occur in Corbino measurements on superconducting thin films, but is much more general both in terms of materials under study as well as probe geometry, i.e.\ it could occur in any frequency-dependent microwave measurement on any non-linear sample. 
In Appendix~C we present data on a quite different material system, VO$_2$ close to its insulator-metal transition near 340\, K, and also there we can identify the characteristic power-dependent oscillations in the spectra.
Also, considering the typical three-standard reflection calibration, it is thus important that the calibration standards do not exhibit non-linearity.

In spectroscopy, one is typically interested in the intrinsic material response and thus wants to avoid extrinsic contributions that affect the measured spectra. 
We thus briefly discuss possible strategies to minimize the unwanted oscillations. One obvious step is reducing standing waves in the coaxial line as much as possible, which will also help for the different calibration schemes. 
But as mentioned, reflection can neither be avoided for the sample nor for the VNA. 
In principle one can suppress standing waves on the coaxial line by increasing the damping, e.g.\ using attenuators, but increased attenuation leads to even stricter requirements for the calibration, which already is the limiting factor for cryogenic Corbino spectroscopy. 
Thus there are no simple hardware perspectives to reduce the oscillations.

Concerning the actual microwave measurements, an obvious strategy is applying as low VNA output power as possible such that the non-linear response does not set in. In many instances this will be possible, in particular if the frequency range of interest is small. 
But if one is interested in wide frequency ranges, e.g.\ up to 20 or even 40~GHz,\cite{Wu1995,Scheffler2005c} then one faces the problem that the attenuation in the microwave line increases with frequency. 
Thus, if one applies the same output power for all frequencies, then the power of the reflected signal reaching the VNA detector will be comparably low for highest frequencies, and thus the output power has to be high enough to allow reliable detection. 
The lower damping at lower frequencies then can lead to much higher power at the sample for low frequencies, which then could induce the non-linear response. 
A possible solution could be an additional amplifier for the reflected signal before being detected by the VNA.

Another approach could be a frequency-dependent VNA output power chosen such that the microwave power at the sample is frequency-independent.\cite{Booth1996b} Such a \lq power flatness\rq{} procedure can be applied for room-temperature testing using VNAs and is implemented by measuring the microwave power at the position of interest using a dedicated power sensor. 
In our case, this strategy is not viable for two reasons: firstly, since damping and phase shift of the microwave line strongly depend on temperature, the power-flatness procedure would be needed for cryogenic temperatures and for each temperature of interest separately. 
Secondly, and more fundamental, the relevant standing wave pattern depends on the sample impedance and thus would not be properly reproduced during a power measurement at the sample position with a dedicated microwave sensor.

Processing the experimental data \textit{a posteriori}, e.g.\ low-pass filtering, could effectively erase the oscillations, but possibly feigning \lq unaffected\rq{} data although heating still influences the filtered spectra.
An elegant and completely different strategy to overcome the oscillations could be using microwave pulses, but pulsed measurements require microwave instrumentation that goes well beyond the established cryogenic Corbino setups.
With no ideal solution at hand, our approach to address this experimental problem will be careful consideration of the applied microwave power and inspection of the measured spectra for possible signatures of this unwanted effect.

\section{Conclusion}

We have shown that experimental microwave spectra of superconducting samples near $T_\textrm{c}$ can contain signatures (\lq oscillations in the spectra\rq) of standing waves in the setup even if a full low-temperature calibration is successfully applied. These oscillations are caused by the combination of non-linear response of the sample and frequency-dependent microwave power at the sample position, with the latter caused by the standing wave pattern of the setup.
This effect is generic to any broadband microwave measurement and non-linear sample, but is particularly relevant for the study of superconductors. Considering the diverse related research activities with microwave spectroscopy, we expect that the \lq standing-wave non-linear oscillations\rq{} have do be taken into account carefully in numerous instances to optimize experimental procedures that then reveal the fundamental material properties of interest.

\section{Acknowledgements}
We thank G.\ Untereiner for preparing the calibration and test samples.
The UNi$_2$Al$_3$ sample was kindly provided by M.\ Jourdan, 
the NbTiN sample by M.\ Quintero-P\'erez and A.\ Geresdi, 
and the VO$_2$ sample by A. Polity.
We thank M.\ Bader, M.\ M.\ Felger, A.\ D'Arnese, M.\ Thiemann, M.\ Javaheri, P.\ Karl, and S.\ Schnierer for support with experiments.
We acknowledge helpful discussions with E.\ Silva and N.\ Pompeo.
Financial support by the DFG is acknowledged.

	
\section*{Appendix A: Samples}

\begin{table}[]
	\begin{tabular}{cccc}
		\hline
		\textbf{sample} & $\begin{array}{cc}\textbf{film}\\\textbf{thickness}\end{array}$ & $\begin{array}{cc}\textbf{substrate}\\\textbf{material}\end{array}$ & $\boldsymbol{T_\mathrm{c}}$\\
		\toprule
		Al & 8\,nm & sapphire & 2.0\,K\\
		\hline
		UNi$_2$Al$_3$ & 150\,nm & YAlO$_3$ & 1.0\,K\\ 
		\hline
		NbTiN & 100\,nm & silicon & \lq 7.9\,K\rq{}\, \cite{TemperatureSetup4}\\
		\hline
		Pb & 60\,nm & glass & 7.3\,K\\
		\hline
		VO$_2$ & 200\,nm & sapphire & $\approx 340$\,K\\
		\hline
	\end{tabular}
	\caption{Overview of the different samples used in this study. $T_\mathrm{c}$ is the critical temperature of the superconducting transition (first four samples) or of the insulator-metal transition (VO$_2$), respectively. The lateral dimensions of all samples are between 4\,mm x 4\,mm and 5\,mm x 5\,mm, and gold contacts in Corbino geometry were applied to all of them.\cite{Scheffler2005a}}
	\label{TableSamples}
\end{table}

To demonstrate how general the discussed effect is and how it can show up for different experimental situations and analyses, we present data obtained on various materials, with the samples listed in Table \ref{TableSamples}. All of them are thin films deposited onto insulating substrates.
The very thin aluminum (Al) film has $T_{\text c} \approx 2.0$~K,\cite{Scheffler2005a} which is substantially higher than $T_{\text c} = 1.2$~K of bulk aluminum, due to its granular structure.\cite{Pracht2016}
UNi$_2$Al$_3$ is a heavy-fermion superconductor with $T_{\text c} = 1.0$~K [\onlinecite{Geibel1991}]; the thin film sample was grown by molecular beam epitaxy on YAlO$_3$\cite{Steinberg2012,Jourdan2004,Scheffler2006} and patterned into stripe geometry before depositing the Corbino contacts.\cite{Scheffler2007}
The 100~nm thick NbTiN film was sputtered on silicon,\cite{vanWoerkom2015,Guel2017} and has actual $T_{\text c}$ substantially higher than suggested by Fig.\ \ref{R_DC-Oszis}.\cite{TemperatureSetup4}

\section*{Appendix B: Interplay of temperature and power}\label{AppendixTemperaturePower}

\begin{figure}
	\centering
	\includegraphics[width=\columnwidth]{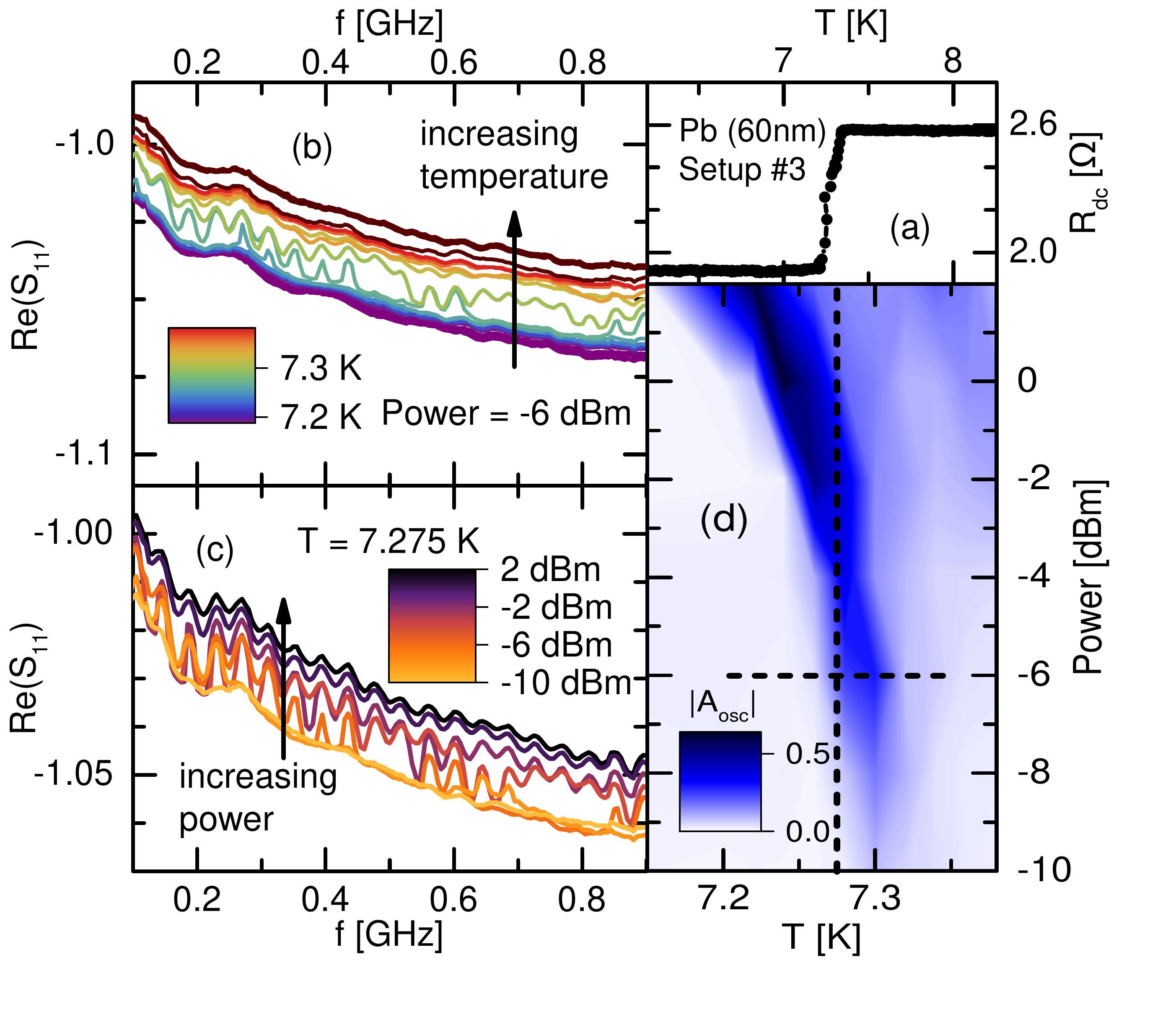}
	\caption{Temperature- and power-dependent microwave measurements of a lead thin film near $T_{\text c}$, using setup \#3.
		\textbf{(a)} Temperature-dependent dc resistance. \textbf{(b)} Spectra of Re($\hat S_{11}$) for several temperatures, at fixed microwave power of -6~dBm. \textbf{(c)} Spectra of Re$(\hat S_{11})$ for different applied microwave powers, at the same nominal temperature 7.275~K. For certain combinations of temperature and power, (b) and (c) exhibit clear oscillations. \textbf{(d)} Fourier transform amplitude $|A_{\text{osc}}|$ for timefrequency $\tau_{\text{osc}}=24$~ns, plotted in color code as function of temperature and microwave power. Dashed lines indicate the tuning of either temperature or power for the spectra shown in (b) and (c).}
	\label{Pb60nm}
\end{figure}

To illustrate the combined roles of sample temperature and applied microwave power, we discuss spectra obtained in setup \#3 on a strip-shaped\cite{Scheffler2007} lead film\cite{Steinberg2010,Ebensperger2016} on glass, with dc data and $T_{\text c} \approx 7.3$~K visible in Fig.\ \ref{Pb60nm}(a).
Microwave spectra obtained near $T_{\text c}$ are shown in Fig.\ \ref{Pb60nm}(b) for several temperatures and in Fig.\ \ref{Pb60nm}(c) for different applied microwave powers. 
The pronounced oscillations are visible in the spectra only for certain combinations of temperature and power. 
We quantify these oscillations by their Fourier transform, which we evaluate at the timefrequency of this dominant oscillation, $\tau_{\text{osc}}=24$~ns, and obtain the corresponding amplitude $|A_{\text{osc}}| = |A_{\text{FT}}(\tau_{\text{FT}}= 24\,\text{ns})|$ of the oscillations. Fig.\ \ref{Pb60nm}(d) plots $|A_{\text{osc}}|$ as a function of temperature and power, and from the clear maximum in these data it is evident that the oscillations only occur in a narrow regime of combinations of temperature (near $T_{\text c}$) and power, with higher powers required for lower temperature. 
(E.g.\ the combinations \{7.30~K, -6~dBm\} and \{7.23~K, 0~dBm\} lead to maximal oscillations for either a given temperature or power.)

\section{Appendix C: Measurements on VO$_2$}
\begin{figure}
	\centering
	\includegraphics[width=\columnwidth]{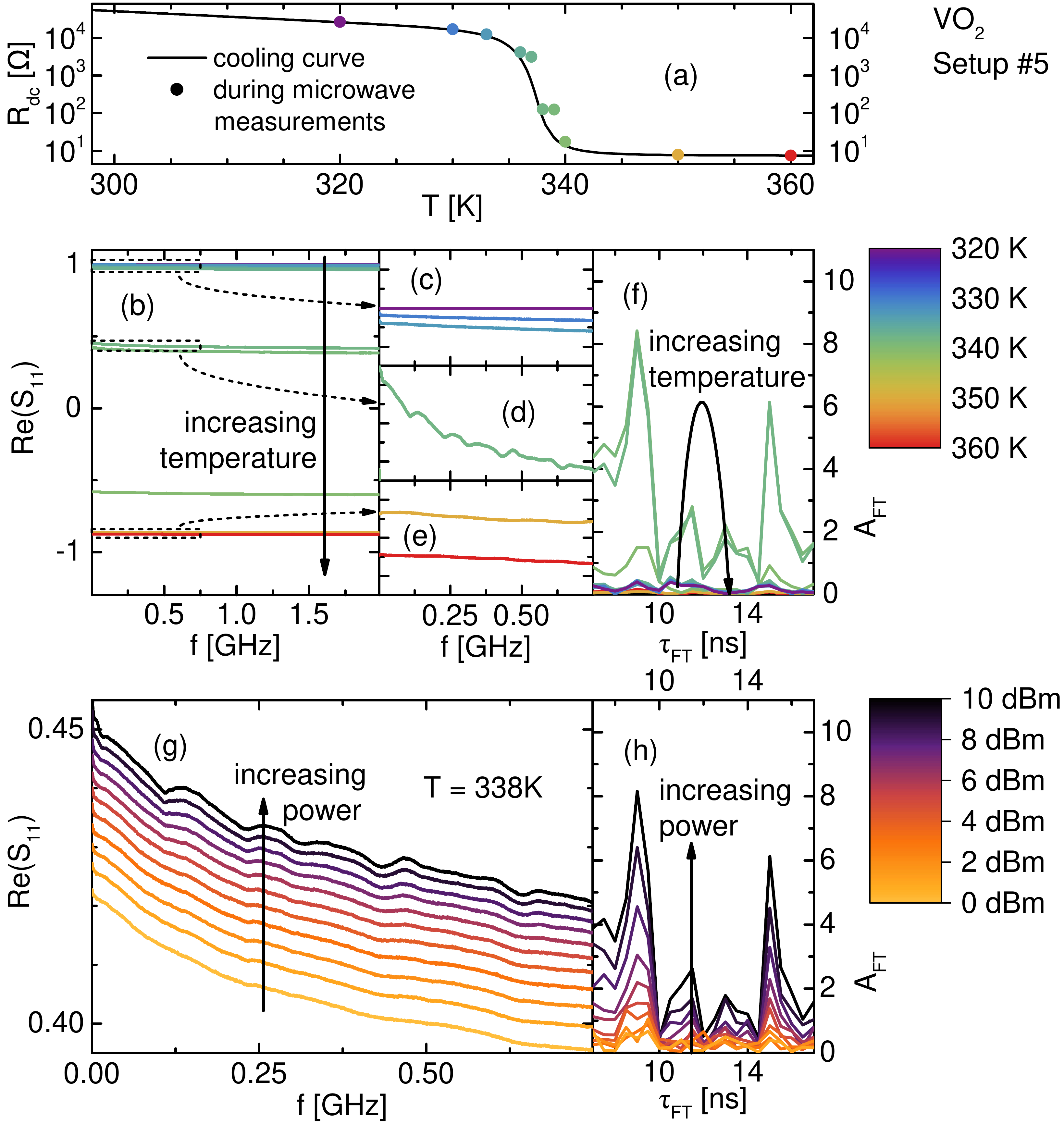}
	\caption{Temperature- and power-dependent microwave measurements on VO$_2$ thin film near the insulator-metal transition temperature (setup \#5).	\textbf{(a)} Dc resistance, where data plotted as circles were obtained simultaneously with the microwave measurements (during setup heating) and plotted in the colors corresponding to the spectra in (b) and data shown as full line were obtained while letting the setup cool down slowly. \textbf{(b)} Spectra of the real part $\text{Re}(\hat S_{11})$ of the reflection coefficient for several temperatures. \textbf{(c)}, \textbf{(d)}, and \textbf{(e)} are magnified plots for the bordered parts of (b), showing data for 320, 330, and 333~K (c), 338~K (d), and 350 and 360~K (e), with the vertical axis of each of the three plots encompassing a 0.03 range. \textbf{(f)} Fourier transforms of spectra in (b). \textbf{(g)} Spectra of $\text{Re}(\hat S_{11})$ at a fixed nominal temperature of 338~K for several applied microwave powers, and \textbf{(h)} Fourier transform of these spectra. Oscillatory behavior in the spectra results in peaks in the Fourier transform.}
	\label{VO2}
\end{figure}

While the superconducting transition might be the most prominent case of a steep $R_\mathrm{dc}(T)$ curve, the above explanation can apply to any material with strong impedance changes as a function of temperature. Therefore, we now consider the very different case of VO$_2$ with its well-studied transition around 340~K from a low-temperature insulating phase to a high-temperature metallic phase.\cite{Imada1998,Dietrich2015}
Here we use setup \#5 to study a 200~nm thick VO$_2$ film grown on sapphire,\cite{Peterseim2016} with the insulator-metal transition evident from the dc data in Fig.\ \ref{VO2}(a).
Microwave spectra of $\text{Re}(\hat S_{11})$ for several temperatures near the transition are shown in Figs.\ \ref{VO2}(b), (c), (d) and (e).
Here, the oscillations of interest are less visible than for the superconducting examples, but the Fourier transform in Fig.\ \ref{VO2}(f) indicates two pronounced peaks at 9~ns and 14~ns, which are strongest for temperatures 337~K to 338~K, consistent with the steepest slope in Fig.\ \ref{VO2}(a). 
Furthermore, when investigating the oscillations as a function of power, shown in Figs.\ \ref{VO2}(g) and (h), we can clearly see that the oscillations strongly increase with power.

The phenomenology and explanation described above for superconductors thus also applies to other materials. But the amplitude of the oscillations for the case of VO$_2$ is much weaker than for the superconducting examples,\cite{CommentTemperatureOscillations} and this can be explained as follows: firstly, the impedance change due to heating is stronger for steeper temperature dependence. 
While the superconducting transitions above have width of order 100~mK, the transition in our VO$_2$ measurement extends over several K.
Secondly, if the standing waves induce an impedance change in a superconductor via heating, then the impedance will increase, leading to even more absorption; i.e.\ the process is self-enhanced.\cite{CommentImpedance} 
For our VO$_2$ case, Fig.\ \ref{VO2}(d) with substantially smaller $|\hat S_{11}|$, this effect will be weaker and even self-inhibiting for somewhat higher temperatures (for $\hat Z$ smaller than $\hat Z_0$ and negative $R(T)$ slope).
Thirdly one has to consider the thermal properties of the materials involved, i.e.\ how much the sample temperature increases for a given absorption of microwave power. 
The rather low specific heat of many materials at cryogenic temperature thus might further enhance the sensitivity of superconductive samples with respect to the standing-wave oscillations.


\begin{thebibliography}{99}
\bibitem{MultiplePasternackCableLengths} Flexible Pasternack Precision Cables, 095 Series PE35611. The used cables have a physical length of 16~cm, 31~cm, 46~cm, and 77~cm, respectively.
\bibitem{TemperatureSetup4}From independent measurements we know that $T_{\text c}$ of the studied NbTiN sample is several K higher than suggested by the data in Fig.\ \ref{R_DC-Oszis}. This is due to the distance between sample and temperature sensor and the large temperature gradients in the probe of setup \#4 when operated at such low temperatures.
\bibitem{fToffset}Manufacturer specifications are for room temperature, and we neglect the temperature dependence of $L_\mathrm{el}$.
\bibitem{CommentTemperatureOscillations}We estimate that the strongest observed oscillations correspond to a temperature swing of 2~mK for the VO$_2$ case and some tens of mK for the superconducting cases.
\bibitem{CommentImpedance} This argument only holds if the sample impedance is lower than the coax characteristic impedance of 50~$\Omega$, and all our cases fall well into this regime.
\end{thebibliography}
\end{document}